\documentclass[preprint, 12pt]{article}

\usepackage{setspace}
\usepackage{url}
\usepackage{amssymb}
\usepackage{graphicx}
\usepackage{natbib}
\usepackage[a4paper, margin=2cm, top=3cm, bottom=3cm]{geometry}
\providecommand{\keywords}[1]{\textbf{\textit{Keywords---}} #1}
\begin{document}
	
	
	\title{A survey on Information Visualization in light of Vision and Cognitive sciences}
	
	\author{BLIND REVIEW}
	\author{Corresponding author: Jose F \mbox{Rodrigues-Jr, junio@icmc.usp.br}\\
	        Luciana A M Zaina+,\\
	        Maria C F de Oliveira,\\
	        Bruno Brandoli,\\
	        and~Agma J M Traina}
	\maketitle
	
	\begin{abstract}
		\noindent{Information Visualization techniques are built on a context with many factors related to both vision and cognition, making it difficult to draw a clear picture of how data visually turns into comprehension. In the intent of promoting a better picture, here, we survey concepts on vision, cognition, and Information Visualization organized in a theorization named Visual Expression Process. Our theorization organizes the basis of visualization techniques with a reduced level of complexity; still, it is complete enough to foster discussions related to design and analytical tasks. Our work introduces the following contributions: (1) a Theoretical compilation of vision, cognition, and Information Visualization; (2) Discussions supported by vast literature; and (3) Reflections on visual-cognitive aspects concerning use and design. We expect our contributions will provide further clarification about how users and designers think about InfoVis, leveraging the potential of systems and techniques.}
	\end{abstract}
	{\ }\\
	\noindent{\keywords{Information Visualization, InfoVis, Vision Science, Cognitive Science}}
	
	\section{Introduction}
	\label{sec:introduction}
	
	\noindent{Understanding why and how visual representations work is an important issue addressed in many works on Information Visualization \citep{DillEtAll2012}\citep{Thomas2005}. On the track of this issue, in this work, we assume that well-designed visualizations must stimulate visual and cognitive processes in a way that reasoning is amplified. The effective promotion of such reasoning depends on principles mastered in the sciences of vision and cognition: vision is the gate through which information derived from computer graphics reaches the brain; cognition refers to the processing that is induced by such graphics. Vision and cognition are closely intertwined, a fact to be considered in the design of visualizations. Accordingly, understanding Information Visualization (InfoVis for short) in light of these sciences may improve design principles that, usually, are performed intuitively. To this end, we review the steps that take place during vision-cognition phenomena when the goal is data analysis. In survey fashion, we compile the literature introducing the following contributions:}
	
	\begin{itemize}
		\item{Theoretical compendium: we draw a descriptive relationship between vision, cognition, and Information Visualization;}
		\item{Discussions: we debate our rationalizations over extensive literature;}
		\item{Reflections: we provide study cases to revisit design practices.}	
	\end{itemize}
	
	
	We organize visualization concepts aiming at the needs pointed out by \cite{JohnsonChallenges06}, who recommend the characterization of how and why visualizations work, and by \cite{Scaife96}, who stress the importance of the cognitive aspects underlying visualizations. Furthermore, we review principles for visual representations that, according to \cite{CardReadings99}, are an initial step towards more effective visualization techniques, a demand defended by \cite{ChungEtAl2012}. 
	
	We note that, while we build an association between Information Visualization, vision, and cognitive sciences, we do not reach a definitive settlement. This is because neither vision nor cognition are yet fully understood. Rather, constrained to the current state of the art, we introduce an organizational process that discusses use and design.
	
	\section{Related work}
	\label{sec:relwork}
	
	\noindent{The literature presents several models that lay the basis to Information Visualization. \cite{Bertin1977} introduced the concept of deriving visual structures from a set of fundamental components. \cite{47}, and \cite{Mackinlay1986}, conducted studies on the usefulness of visual patterns in the form of frameworks for design. \cite{CardReadings99} follow Bertin by discussing the importance of the spatial substrate. \cite{14} suggests a taxonomical space for quick referencing; \cite{29} reasons about the possibilities of visualization and interaction; and \cite{ChiTaxonomy00} focuses on data transformations. In the realm of design, \cite{BugajskaThesis03} deals with spatial and abstract visualizations considering the design guidelines of \cite{TweedieExternalizations97}. In the line of works that reflect about the visualization field, \cite{Wijk2006} systematically discusses visualizations based on a cost-oriented analysis; and \cite{Green2009} research many cognitive and perceptual aspects \citep{Rensink2000} to build a model and a set of guidelines for design. \cite{Patterson201442} introduces a framework based on vision and cognition sciences, similar to ours, focusing on top-down processes. Comparatively, we conduct a complementary bottom-up approach; we translate vision-cognition phenomena into an original vocabulary that may bring such science closer to the design practice.}
	
	To organize our compilation on vision and cognitive sciences, we depart from the Visualization Pipeline of \cite{CardReadings99} to draw the Visual Expression Process, a sequence of events delineated by the possibilities of the visual-cognitive interplay. According to our organization, (1) vision phenomena (pre-attentive stimuli) determines a map of potential interesting objects. Then, attentive selection concentrates on one single element, part of a set of (2) analytical perceptions. Such perceptions go through a pattern-matching process that turns them into (3) abstract patterns that, in working memory, support cognition in combination with domain knowledge originated from long-term memory. Finally, leading to (4) cognitive decision support. Nevertheless, our process does not explain the intercourse between vision, cognition, and Information Visualization; this is not feasible considering the current knowledge, and neither it would fit in a single article.
	
	The rest of the paper is organized as follows. Section \ref{sec:vision_and_visualization} draws a detailed panorama of vision and cognition in the realm of Information Visualization. Section \ref{sec:visual_expressivity} introduces the Visual Expression Process, our organizational theorization; and the last section presents conclusive remarks.\\
	
	\section{Concepts on Vision, Cognition, and Visualization}
	\label{sec:vision_and_visualization}
	
	\noindent{According to Vision Science, the visual process has two stages, namely, the parallel extraction of low-level properties, called {\it pre-attentive processing}, followed by a slower {\it detailed scan}. The first stage promotes the major benefit of visualizations, that is, improved data comprehension \citep{43}. Meanwhile, the second stage addresses conventional reading practices that do not contribute towards faster cognition, but that are necessary for further analysis. In fact, \cite{WarePerception04} states that understanding what is processed pre-attentively is probably the most important contribution that Vision Science can bring to visualization.}
	
	This two-stage process might be the underlying motivation for the broadly referenced ``visualization mantra'': overview first, zoom and filter, then details on demand \citep{29} -- Schneiderman came to this conclusion by means of intuition and empirical experiments, never drawing the vision/cognitive reasons of why this is the case. We discuss this process in the following sections according to our process, presented in section \ref{sec:visual_expressivity}.
	
	
	\subsection{Maps of Saliences}
	\label{subsec:saliency_map}
	
\begin{figure*}[htb]
		\centering
		\includegraphics[width=0.8\textwidth]{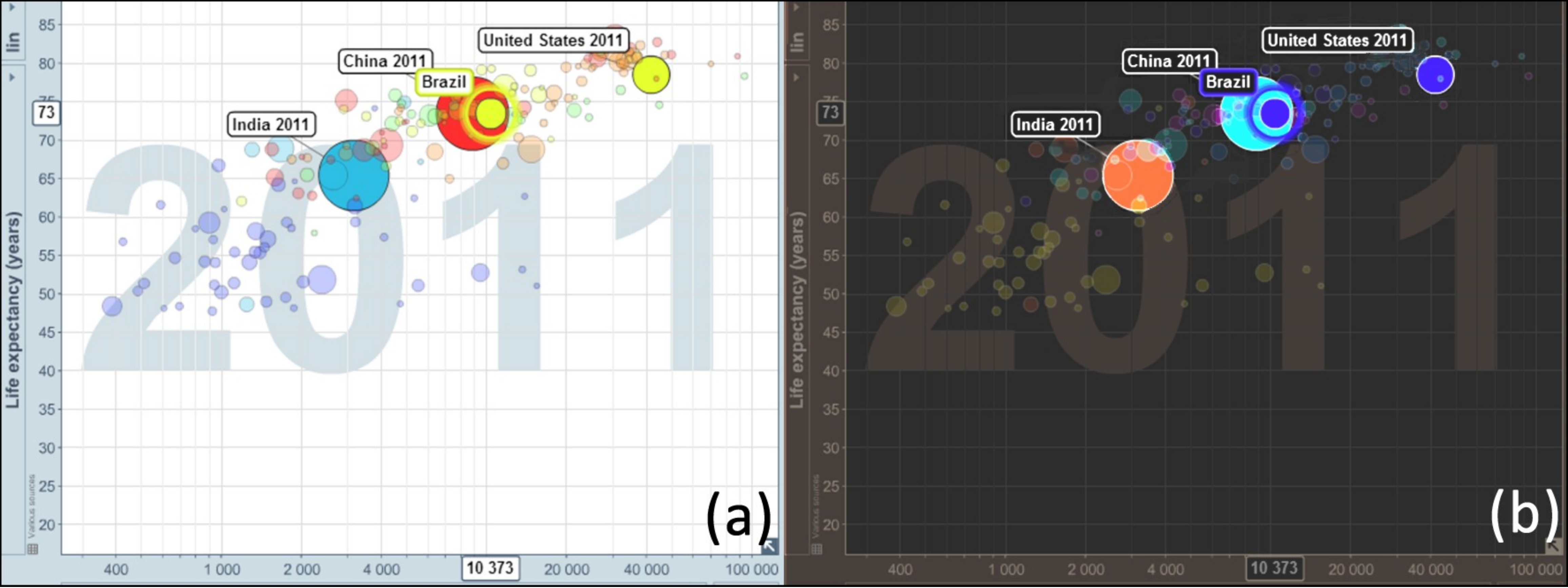}
		\caption{(a) Example of a map of saliences over the Gapminder tool. In the scene, the bubbles with outstanding features stand for saliences corresponding to countries of interest. (b) The result is a map of potential targets that pop out due to their color, or their shape.}
		\label{fig:saliency_map}
	\end{figure*}

	\noindent{In the early stages of vision, the brain deals with the problem of casting potential elements of interest; regions of the scene that should be considered for cognition. To cope with that, a complex process takes place so that certain characteristics pop out to the eyes. These characteristics appear as saliences over the scene. They define the so-called {\it maps of saliences}, exemplified in Figure \ref{fig:saliency_map}; the first element of our theoretical organization.}
	
	The principle of salience is to reinforce the perception of the areas in the scene whose visual properties contrast with those of their surroundings \citep{Itti2001}. It is a process that considers different visual aspects, such as position and color, and that ends as it combines them onto a single scene \citep{Nothdurft2000}.
	
	The occurrence of salient features stems from their interplay with other stimuli, depending on a context that favors conspicuousness. The brain is highly trained to detect such configurations, being able to track salient features in parallel, in real-time, and covering the entire visual field. There is also evidence that this ability is strongly influenced by the task-at-hand, in a top-down (cognition-to-vision) process \citep{Navalpakkam2007} \citep{Patterson201442}.
	
	\subsection{Attention and Attentive Selection}
	\label{subsec:attention}
	
	\noindent{Although the brain perceives visual targets simultaneously in a map of saliences, it cannot process all of them in parallel. This is considered a prohibitively computational task even to the most sophisticated brains \citep{Tsotsos1991}. Primates and other animals handle this by restricting the consideration of the objects presented to their eyes: their vision concentrates on small regions considering objects one after the other. This is a serialization process ruled by what is called {\it attention}. In other words, once a map of candidates is ready, it is necessary to ``filter out'' one of these candidates for attention. Attention, here, occurs in accordance with the task that the user is performing; the task determines the information demands and, consequently, what visual stimuli should be extracted from the scene. Trial and error is part of the process, which is iterative.} 
	
\begin{figure*}[htb]
		\centering
		\includegraphics[width=0.8\textwidth]{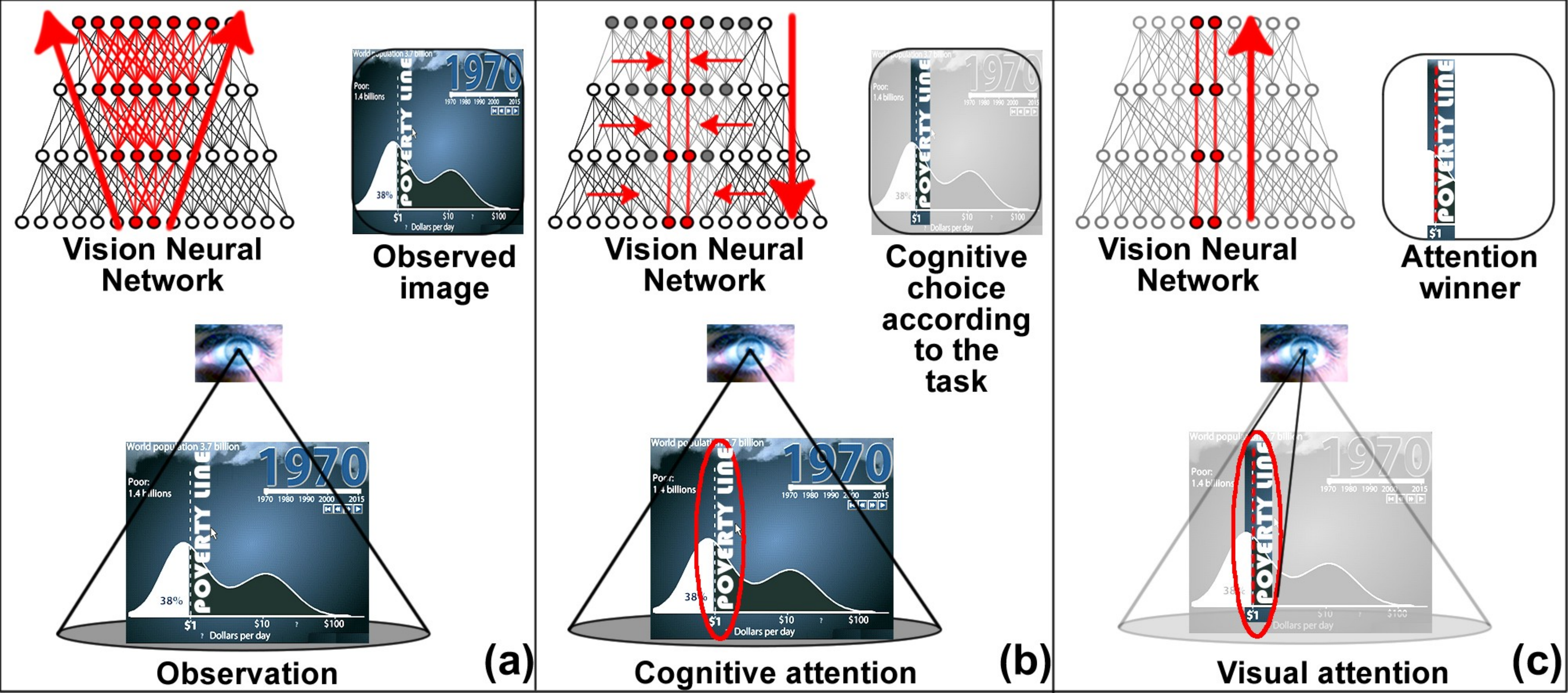}
		\caption{Three-stage pyramidal visual selection: (a) bottom-up feed-forward, (b) top-down winner-take-all feedback \citep{Lippmann1987}, and (c) bottom-up straight path.}
		\label{fig:attentive_selection}
	\end{figure*}

	This process of filtering has been modeled as a pyramidal neuronal structure -- named {\it selective tuning} model \citep{Cutzu2003} \citep{Essen1992}. This theoretical model predicts a broad layer of neurons in its first level, narrowing down as it advances to upper layers. The layers intercommunicate through feed-forward and feedback connections and, according to \cite{Cutzu2003} \citep{Tsotsos1995}, a pyramid of neurons successively performs three stages of processing, illustrated in Figure \ref{fig:attentive_selection}: (a) bottom-up feed-forward, (b) top-down winner-take-all feedback \citep{Lippmann1987}, and (c) bottom-up straight path. This process explains what is broadly referenced as {\it filtering} or {\it attentive selection}.
	
	Following this widely-accepted model, it is worthy to note that, as described, filtering occurs according to both bottom-up and top-down processes. The bottom-up process depends on the visual stimuli, while the top-down process depends on the task at hand and is prioritized \citep{Wolfe94}. Each process influences the other iteratively. Actually, according to \cite{JSID:JSID1939}, the entire visual-cognition interplay in influenced by top-down mechanisms that emanate from working memory, but that may be highly influenced by long-term memory \citep{woodman2013we}. In this work, we focus on the bottom-up process, but there are works that explore higher-level factors related to the top-down process \citep{Patterson201442}.
	
	\subsection{Cognition, Memory and Vision}
	\label{subsec:vision_and_memory}
	
	\noindent{After a target is selected, it is potentially useful for ``details on demand'', or cognition. In general terms, cognition refers to the acquisition or use of knowledge \cite{cohen85}, and may occur analytically (conscious and slower) or intuitively (automatic and fast), as defended by the dual system theory \citep{Evans2008}\citep{Evans01052013}. Cognition is of special importance in data visualization as it supports analogical reasoning \citep{Patterson2009}. That is, the transfer of inferences from a relationship of elements in one domain (the analogue) to a relationship of elements in another domain (the target). In any case, cognition is intermediated by the memory system. The relationship between memory and cognition is studied by works on Cognitive Architectures, such as ACT-R \citep{Buttner2010} and Soar \citep{Young1999}.}
	
	In Soar and other theories, the {\it structural configuration} of memory roughly reflects the model of \cite{Baddeley1974}. Working memory includes three components: the central executive module, the phonological loop, and the sketchpad. The central executive module determines the attention focus, guiding the visual system, for example, by top-down biasing the pyramidal selection mechanism. The phonological loop stores information related to sound. The sketchpad (also known as Visual Short-Term Memory -- VSTM \footnote{although, this is not a consensus}) is associated with the maps of saliences discussed in Section \ref{subsec:saliency_map}, storing information related to space and to visual features.
	
	Following these lines, memory comes to be the main element in supporting cognition as it allows complex mental operations. Consequently, it also supports the practice of data visualization. Miller's Law \citep{Miller56} states that the memory is limited to 7+/-2 elements; more recent, and accepted, works state that this limit is at 4 {\it chunks} of elements -- a chunk refers to the grouping of elements into larger units based on their meaning \citep{Cowan2010}. \cite{alvarez2004capacity} discuss memory limits considering the nature of the elements to be remembered, demonstrating that, despite the consensus of a (very) limited resource, there is not an ultimate conclusion about the topic.	In any case, these limitations are severe because the greater the capacity of an individual's memory, the more information she/he has available for solving problems \citep{Just1992}.
	
	The memory system supports cognition in two ways: by retaining a list of elements for quick referencing, and by assisting in the construction of {\it mental models} \citep{Laird83}. According to \cite{johnson2010mental}, mental models preserve the relationship between entities by defining analogies that save on logical reasoning, one of the principles behind complex visualization techniques and interaction \citep{Liu:2010:MMV:1907651.1907983}. Following the study of \cite{Logie95}, mental models are created in the visuo-spatial sketchpad, a specialization of VSTM. Stimuli encoded into VSTM, ruled by cognition, may activate enduring information known as long-term memory, including facts, meanings, relationships, skills, and procedures \citep{patterson2010implicit} -- for a deeper discussion on working memory and long-term memory, refer to the work of \cite{rose2010similarities}.
	
	The concepts presented in the former sections rely on ideas posed by widely accepted theories, among several others, for the visual-cognitive process. The choice for this specific line of thought has been motivated by its intuitive coherence and scope of influence in the literature. Notwithstanding, other theories are widely referenced, such as the Spotlight \citep{Eriksen1973} and the Gradient \citep{Cheal1994} models. For an ample discussion, refer to the work of \cite{squire2009memory}.
	
	\section{The Visual Expression Process}
	\label{sec:visual_expressivity}
	
\begin{figure*}[htb]
		\centering
		\includegraphics[width=\textwidth]{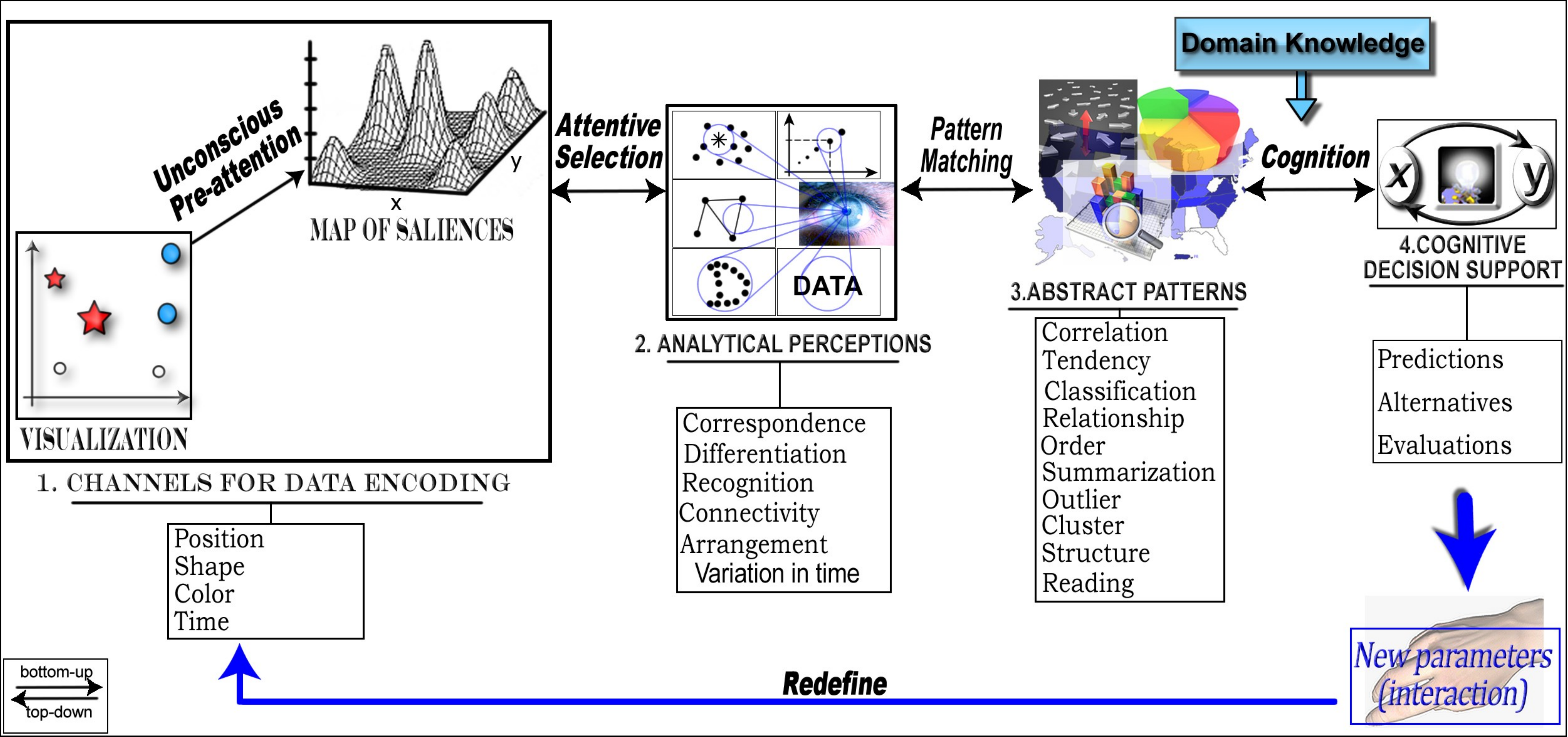}
		\caption{The Visual Expression Process for Information Visualization specified with four constituents. The black bidirectional arrows refer to bottom-up/top-down visual-cognitive processes. In blue, new parameters of interaction alter the scene iteratively.}
		\label{fig:VisualExpressionProcess}
	\end{figure*}

	\noindent{In this section, we review the practice of visualizing data by considering the concepts presented so far. We organize the relationship among visualization, vision, and cognition according to a framework named Visual Expression Process -- Figure \ref{fig:VisualExpressionProcess}. Our theoretical organization has four constituents: (1) pre-attentive stimuli, (2) analytical perceptions, (3) abstract patterns, and (4) decision support. For completeness of our survey, Table \ref{tab:models} summarizes previous models found in the literature. Our theorization is inspired by these previous works putting together concepts in a complementary point of view.}
	
		\hyphenpenalty=10000
		\begin{table*}[htb]
			\begin{center}
				\caption{Summary of previous models on InfoVis.}
				\label{tab:models}
			\end{center}
			\begin{tabular}{|p{1.01in}|p{3.2in}|}	
				\hline
				{\footnotesize{\it{Model}}}
				&{\footnotesize{\it{Summary}}}\\
				\hline
				\hline
			\end{tabular}
			
			\begin{tabular}{|p{1.01in}|p{3.2in}|}
				{\footnotesize{\cite{Lohse:1993:CMU:1461776.1461779}}}
				&{\footnotesize{An algebraic model to estimate the effort to answer questions based on a visualization.}}\\				
				\hline
				{\footnotesize{\cite{CardReadings99}}}
				&{\footnotesize{A pipeline of how to conduct the visualization practice.}}\\
				\hline
				{\footnotesize{\cite{1532781}}}
				&{\footnotesize{An economic model stated to evaluate efficacy and efficiency.}}\\
				\hline
				{\footnotesize{\citep{Ware2005}}}
				&{\footnotesize{A top-down (problem-solving) model that states that what we see in a visualization depends on what we are seeking for.}}\\
				\hline
				{\footnotesize{\citep{TOPS:TOPS1150}}}
				&{\footnotesize{A descriptive discussion of how visualizations work that culminates into a compilation of principles and perspectives.}}\\
				\hline
				{\footnotesize{\citep{Patterson201442}}}
				&{\footnotesize{A vision-cognition model that explains visualization and how to improve it from a top-down perspective.}}\\
				\hline
			\end{tabular}
		\end{table*}
		\hyphenpenalty=50

	According to our theorization, (1) pre-attentive stimuli come from the neuronal reaction to light, determining a map of potential interesting objects, or saliences. Then, attention concentrates on one single element, part of a limited taxonomic vocabulary of (2) analytical perceptions. Such perceptions go through a pattern-matching process that turns them into (3) abstract patterns. Finally, abstract patterns in working memory support cognition in combination with domain knowledge originated from long-term memory, leading to (4) decision support. Notice, in the figure, that the arrow in between each pair of the four constituents is bidirectional to indicate that the interplay occurs both bottom-up and top-down, notably in response to the task at hand. Also, notice that after cognition, there might be new information demands, what leads to the redefinition of the visualization by means of new parameters for interaction -- as depicted in blue in Figure \ref{fig:VisualExpressionProcess}. Following, we present further details.
	
	
	\subsection{Pre-attentive Stimuli - channels for data encoding}
	\label{subsec:pre_attentive_stimuli}
	
	\noindent{Pre-attentive stimuli impel maps of saliences, as highlighted on the leftmost side of Figure \ref{fig:VisualExpressionProcess}. In the work of \cite{RodriguesDesignSpace08}, the authors verified that such stimuli manifest through position, shape, color, and time. Here, we refer to these factors as {\it channels for data encoding}, in the sense that they encode data into visual stimuli.}
	
	Although the consideration of four channels is a reductionist classification, it is supported by the literature. About {\em color and texture}, \cite{Watt1995} affirms that, just like texture, color is the psychological response to the spectral characteristics of a surface; and that, different surfaces are perceived as having different colors. Furthermore, \cite{Motter1994} observes that, early in visual processing, the incoming information is sorted and grouped according to the similarity of simple {\it shape features}, such as {\em orientation} or {\em size}, and of surface features, such as {\em color}, {\em luminance}, or {\em texture}.
	
	The features of each channel span to a large set, but \cite{CardReadings99} observe that just a limited number of the many existing graphical properties are used for Information Visualization. \cite{healey2012attention} present an extensive survey on pre-attentive features and visualization. A non-exhaustive list of such features is presented in Table \ref{table:pre-attentive_features}.

{\centering
	\begin{table*}[htb]
		\caption{Classes/channels of pre-attentive features from the perspective of data representation.}
		\label{table:pre-attentive_features}
		\begin{tabular}{p{3.5cm} p{8.35cm}}
			\hline  
			{\footnotesize{\it{Pre-attentive class (\mbox{data encoding channel})}}}
			&{\footnotesize{\it{Features}}}\\
			\hline
			
			{\footnotesize{Position}}
			&{\footnotesize{1D/2D/3D position, stereoscopic depth;}}\\
			
			{\footnotesize{Shape}}
			&{\footnotesize{line, area, volume, form, orientation, length, width, collinearity, size, curvature, marks, numerosity, convex/concave;}}\\
			
			{\footnotesize{Color}}
			&{\footnotesize{hue, saturation, brightness and texture;}}\\
			
			{\footnotesize{Time}}
			&{\footnotesize{movement, morphing, blinking, color/light intermittence.}}\\     \hline
		\end{tabular}  
	\end{table*}
}
	
	\subsubsection*{Discussion}
	A well-designed visual representation must present a high overlap between its map of saliences and its (implicit) map of semantic relevance. That is, the design of visual representations is supposed to maximize pre-attentive effects. However, such maximization may not be possible without flexible human intervention over the channels of data encoding. In other words, interaction, the active redefinition of the channels, is a mandatory need not fully explored in many designs.
	
	 \cite{LiuStasko2008} point out that many visualization systems are not sufficiently flexible to support user customization and appropriation. In fact, it is not difficult to find visualization tools that are limited in allowing the user to determine how to encode data. In these circumstances, a user may ask ``may I change the positioning order of the elements?'', ``can I have each year represented with a different shape?'', ``can I color the left group in red?'', or ``can I see that animated?''. Each of these examples refers to a particular pre-attentive feature or, as we propose, to a data encoding channel.
	 
	 For instance, consider the seminal system GGobi \citep{CookSwayne2007}, which introduced a large set of features if compared to its former version, system XGobi. Still, a brief analysis reveals that there is much to be improved: positioning of views is limited, except for Parallel Coordinates; shape is not an option for coding in the same way that color is; and animation is restricted to scatter plots through touring techniques. Those design issues contrast to what is observed in commercial systems like TIBCO's Spotfire  (\url{http://spotfire.tibco.com/}) and Google's Gapminder (\url{http://www.gapminder.org/}), which present higher levels of freedom for each encoding channel, but that, still, do not achieve full appropriation of the scene the way we discuss.	
	
	Accordingly, the design of visualizations must not only allow users to alter visual attributes, they must incorporate mechanisms to emphasize each channel. This is necessary because a given stimulus (position, shape, color, or time) may fail to capture attention when it is surrounded by other stimuli -- this is called inattentional blindness. In such circumstance, this given stimulus fails in competing with other more interesting stimuli or, worse, it is ignored \citep{Mack1998}\citep{Simons2000}. ``Blindness'' occurs especially when the viewer has some previous expectation on the scene; in such situations, top-down processes can strongly influence on what is noticed. This issue has been noticed by \cite{Patterson201442}, who claim that the design of a visualization technique must provide means to capture attention alerting users of changes. 
	
\begin{figure*}[htb]
		\centering
		\includegraphics[width=0.8\textwidth]{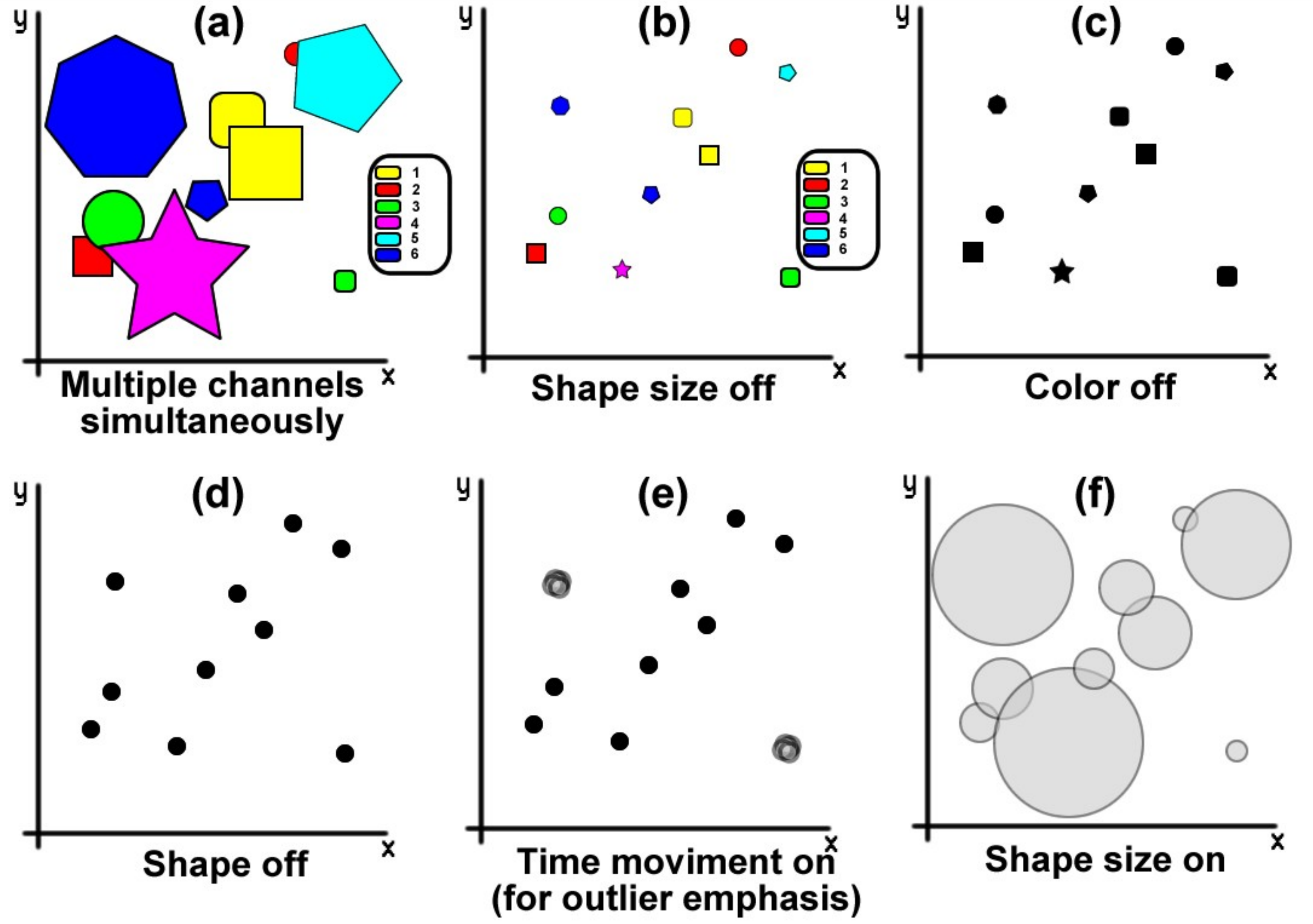}
		\caption{(a) Dispersion plot using channels of position, shape contour, shape size, and color. (b) Shape size turned off. (c) Color turned off. (d) Shape contour turned off. (e) Time movement turned on for automatically detected outliers. (f) Shape size turned on.}
		\label{fig:channelstogle}
	\end{figure*}

	Furthermore, according to the limitations of the memory system, as presented in Section \ref{subsec:vision_and_memory}, memory overloading is a problem that might constrain problem-solving. In fact, according to \cite{Wickens01062008}, multiple tasks in one same cognitive dimension might decrease performance. One possibility to lessen this drawback is to selectively turn the visual channels on and off. For example, it is possible to conceive a dispersion plot in which size is simply turned off, so to temporarily avoid overlapping -- see Figure \ref{fig:channelstogle} (b). Similarly, it is possible to turn off color, presenting all the information with black marks, emphasizing the role shape contours -- Figure \ref{fig:channelstogle}(c). It is also possible to include intermittence, or movement, much the way that cyclists do with lights in urban traffic. This way, specific graphical elements can start to blink -- Figure \ref{fig:channelstogle}(e). We conjecture that this kind of selective constraining of channels not only alters attention mechanisms, as there will be fewer candidates for attention, it also reduces the memory load, as there would be fewer targets. Furthermore, it has the potential of affecting top-down processes by de-emphasizing features that are expected by the user -- and that prevent her/him from noticing unexpected traits due to ``blindness''. In these cases, algorithms to assist visualizations may be helpful, as advocated by \cite{Chaomei2006}. Such algorithms might monitor statistical features or mine characteristics of interest to suggest new encodings ``on the fly'', enabling more intense appropriation of the scene. However, not much research has been conducted on this issue, which has potential for further developments.
	
	\subsection{Analytical Perceptions}
	\label{subsec:visual_perceptions}
	
	\noindent{Data encoding channels provide maps of potential targets for attention. Now, following vision theory, the next mechanism is attentive selection -- see Figure \ref{fig:VisualExpressionProcess}. Biased by user intention, a subset of the prominent entities in a visualization will reach {\it working memory}. Once selected, the chosen visual stimuli will be the basis of the analogies that lead to mental models, see Section \ref{subsec:vision_and_memory}. Here, one question comes up -- what perceptions are produced by the targets of attention in an information visualization design?}
	
	To answer this question, we have extensively inspected the literature tracking the ways in which visual manifestation occurs when the intent is data analysis. We have found a limited set of possibilities, defining a {\it visual taxonomic vocabulary} whose elements appear recurrently. We refer to these elements {\it analytical perceptions}, depicted in the second part of Figure \ref{fig:VisualExpressionProcess}. In the realm of interaction, \cite{yi2007toward} followed a similar procedure based on extensive literature inspection; they achieved results concerning the user intent, producing a limited set of recurrent possibilities.
	
	Analytical perceptions are the traits that any user attentively seeks for in a visual representation. Following the dual system theory presented in Section \ref{subsec:vision_and_memory}, such perceptions occur intuitively. Our investigation indicates that such elements include {\em correspondence}, {\em differentiation}, {\em recognition}, {\em connectivity}, {\em arrangement}, and {\em variation in time}. The most verified of these phenomena, {\em correspondence} and {\em differentiation}, are noted by \cite{Bertin1977} and by \cite{CardReadings99}. The third analytical perception is presented by \cite{Mackinlay1986} who states that the notion of relationship among graphical entities comes from the perception of {\em connectivity}. Meanwhile, {\em arrangement} arises from group positional configurations, largely studied by the Gestalt psychology \citep{KoffkaGestalt35} as observed, for example, in graph layouts \citep{Dwyer2009}\citep{IoannisGDrawing98}. {\em Recognition}, in turn, takes place as a resemblance to previous knowledge and/or expertise, a concept studied in psychological models and Information Visualization models. As \cite{Liu2010} point out, the concept of internalization involves the encoding of perceived information into long-term memory (enduring information or pattern). Lastly, {\it variation}, manifests only along time -- not necessarily for temporal data -- and in combination to the other five perceptions.
	
	
	The notion of analytical perceptions becomes evident when they are not found and the pipeline outlined in Figure \ref{fig:VisualExpressionProcess} is broken, preventing Visual Expression -- if none of the aforementioned perceptions occur, the user is unable to make sense. As depicted in our theorization, analytical perceptions occur after pre-attention (Section \ref{subsec:pre_attentive_stimuli}) and before abstract patterns (Section \ref{subsec:interpretations}), independently of the data domain. Hence, they bridge vision and data interpretation. Specifically, we discuss the analytical perceptions and how they relate to the data encoding channels in the following.
	
	\renewcommand{\labelitemi}{$\bullet$}
	\begin{itemize}
		\item{\em{correspondence}}: each position/shape/color has a direct correspondence to a referential map -- discrete or continuous -- that is part of the scene (explicit) or that is mental (implicit), defining analogical reasoning as explained in Section \ref{subsec:vision_and_memory}. Explicit maps include axes, geographical maps, shape/color dictionaries, and position/shape/color ranges. Implicit maps include known orderings and shape metaphors;
		\item{\em{differentiation}}: each position/shape/color discriminates graphical items. Differentiation is a correspondence achieved by the user, who creates a referential map in memory. Such map is limited in the number of elements (or differentiations) according to the limitations of memory;
		\item{\em{recognition}}: positions/shapes/colors whose decoding comes from the expertise of the user or from previous knowledge -- recognition is a correspondence established from visual entities to concepts retained/learned in long-term memory;
		\item{\em{connectivity}}: shapes, mainly edges, that convey information about relationships among entities in memory;
		\item{\em{arrangement}}: Gestalt principles of organization -- positional placements (closure, proximity, and symmetry) that convey perception about group properties, for example, clusters and structural cues;
		\item{\em{variation in time}}: obtained when the parameters of position/shape/color are altered along time, inducing new perceptions for each of these channels.\end{itemize}
	
	\subsubsection*{Discussion}	
	Our set of analytical perceptions is not an exhaustive listing, but a first reference that suggests that visualization designs tend to resort to the same basic set. For instance, the design of Google's Gapminder tool, although contemporaneous, reproduces the same dispersion plots of statistical books a hundred years old. Nevertheless, as observed by \cite{LiuEtAl2014}, works on new designs and on the evaluation of existing ones have not considered that there is a limited set of elements to instantiate in data representations. This fact could fruitfully support the definition of design languages and frameworks, which would benefit from recurrent constructs that lead to a limited set of analytical perceptions. Differently, current languages and frameworks rely on graphical patterns and it is up to the user to build the desired analytical perception; see Table \ref{tab:language} for a representative set.
		
	\hyphenpenalty=10000
	\begin{table*}[htb]
		\begin{center}
			\caption{Previous works on languages and frameworks for visualization design.}
			\label{tab:language}
		\end{center}
		\begin{tabular}{|p{1.01in}|p{1.0in}|p{2.2in}|}	
			\hline
			{\footnotesize{\it{Work}}}
			&{\footnotesize{\it{Approach}}}
			&{\footnotesize{\it{Elements}}}\\
			\hline
			\hline
		\end{tabular}
		
		\begin{tabular}{|p{1.01in}|p{1.0in}|p{2.2in}|}	
			{\footnotesize{Protovis \citep{BostockProtovis}\citep{Heer2010}}}
			&{\footnotesize{Graphical}}
			&{\footnotesize{Marks, shapes and layout}}\\
			\hline
			{\footnotesize{$D^3$ \citep{BostockD3}}}
			&{\footnotesize{Visualization pipeline}}
			&{\footnotesize{Selection, operation, join, layout and transformation}}\\
			\hline
			{\footnotesize{Improvise \citep{1382904}}}
			&{\footnotesize{Link and coordinate}}
			&{\footnotesize{Variable, function and view}}\\
			\hline
			{\footnotesize{Prefuse \citep{HeerPrefuse}}}
			&{\footnotesize{High-level API}}
			&{\footnotesize{Filter, layout, interaction, color and size}}\\
			\hline
			{\footnotesize{ggplot2 \citep{ggplot2}}}
			&{\footnotesize{Domain specific}}
			&{\footnotesize{Layer, scale, coordinate system and facet}}\\
			\hline
			{\footnotesize{Flexible Linked Axes \citep{6064997}}}
			&{\footnotesize{Linked axes}}
			&{\footnotesize{Axes mapping, interaction, line and point}}\\
			\hline
			{\footnotesize{This work}}
			&{\footnotesize{Visual-cognitive}}
			&{\footnotesize{Pre-attentive encoding channels and analytical perceptions}}\\
			\hline
		\end{tabular}
	\end{table*}
	\hyphenpenalty=50
	
	By considering the concepts reviewed so far, it is possible to conceive a design language whose approach is based on cognition, and whose elements are interactive encoding channels, and analytical perceptions -- refer to Table \ref{tab:language} for comparison. For example, in this design language, one would be able to state a visualization by choosing channel {\it color} and perception of {\it differentiation}. This same visualization would demand a few more elements, as channel {\it position} and perception of {\it correspondence}. As in any language, these elements would receive parameters according to an extensible library of data-to-marks mapping. In the case of relationships, one would be able to choose channel {\it position} with perception of {\it arrangement}, together with channel {\it shape} and perception of {\it connectivity}. Underlying algorithms would abstract undesired complexity, as an algorithm for outlier detection to support color differentiation; and a force-directed algorithm to support arrangement. As a design language, this approach would bring the benefit of discriminating the recurring elements of visualization techniques and having them in libraries for composition. This alternative approach contrasts with the usual practice of combining recurrent elements in ensembles assumed as new techniques. 
	
	
	\subsection{Abstract Patterns}
	\label{subsec:interpretations}
	
	\noindent{According to \cite{Hutchins96}, tools -- or externalizations \citep{Hegarty2004} --  transform difficult tasks into in-mind manipulations of physical systems, or into pattern-matching problems \citep{Giere2002}. Pattern matching, or pattern-recognition \citep{Patterson201442}, the association of a given stimulus to information retrieved from memory \citep{Eysenck2003}, is one of the principles of data visualization and the second step of the Visual Expression Process  -- see Figure \ref{fig:VisualExpressionProcess}.}
	
	According to our organization, once a user focuses on an analytical perception, she/he proceeds to match this perception to an abstract pattern. Based on the theory of vision -- seen in Section \ref{subsec:vision_and_memory}, the generation of patterns is supported by memory, which is filled with data from the visual-sensorial system or from long-term memory. The efficiency of this intercourse is explained by the fact that the visual-sensorial system provides spatial information to memory at rates higher than that of long-term memory \citep{WarePerception04}, in a time ranging from $100$ to $250$ $ms$ \citep{Kieras1997}. Hence, analytical perceptions in the visual-sensorial system work similarly to the images stored in long-term memory, providing efficient pattern-matching. According to the dual system theory, Section \ref{subsec:vision_and_memory}, this process is conscious and slow; but, with practice and experience, a given analyst can become a {\it specialist}. For specialists, pattern matching becomes intuitive \citep{Evans2008} -- refer to the work of \cite{patterson2010implicit} for a thorough discussion. It is a straight conclusion, then, that a period of practice and experience is necessary for most visualization techniques, and that this is an obstacle for use.
	
	A suggestive set of the patterns -- third part of Figure \ref{fig:VisualExpressionProcess} -- that arise from the perception-to-pattern matching, includes: correlation, tendency, classification, relationship, order, summarization, outlier, cluster, structure, and reading. \cite{Tufte1997} and \cite{Amar:2005:LCA:1106328.1106582} provide more exhaustive listings that follow different rationalizations.

\subsubsection*{Discussion}	
	In our literature review, we noticed that many systems disregard the fact that visualization techniques converge through abstract patterns. According to current practices, instead of a set of familiar patterns, the user has to choose among a set of visualization designs that, quite often, they have little experience with \citep{Chen2005}. \cite{Thomas2005} reinforce this notion; they state that abstract patterns correspond to the second factor of their four-steps analytical-reasoning process. Following their process -- a pattern-to-construct sequence, users have constraints in relation to what they can search for in face of a given pattern. For example, suspiciousness tends to appear by means of tracking for outliers; while evidence of illegal lobbying practices emerge from clusters; and community detection in graphs is a task for relationship. Still, users are often offered a menu whose options are, for instance, dimensional stacking \citep{146386}, star coordinates \citep{Kandogan00}, and table lens \citep{24}; alternatives far from the pattern-to-construct task they have in their minds. A similar notion dates back to the 1990's, in the work of \cite{146375}, who advocates in favor of more specific problem-oriented choices. We suspect that, because design-oriented interfaces neglect the more natural notion of abstract patterns, this is possibly one of the reasons why advanced visualization techniques have struggled to achieve a wider commercial dissemination, e.g, in office suites and in everyday spreadsheets.
	
	Take, for instance, the visualization technique Treemap \citep{ShneidermanTreemap92}, introduced for visualizing hierarchical data in general. In two decades, Treemap gained popularity at the academy, but, as a general hierarchical tool, it has failed in reaching a wider use. This is, possibly, because it has been criticized since its introduction \citep{Barlow2001, Cawthon2007, Fabrikant2005}, being accused of lacking cognitive plausibility, having poorly perceived aesthetic qualities, and presenting poor task-driven performance \citep{Wood2008}. Despite all, however, an especial design of the Treemap has remarkably succeeded. The SequoiaView system (\url{www.win.tue.nl/sequoiaview}) has achieved wide dissemination (check \citep{Wijk2006} for some impressive numbers), far beyond the academic walls. But, how can we have two flavors of the same technique evolve in different ways? Certainly, not one single aspect explains everything, but an outstanding factor comes up: the SequoiaView is domain and pattern-oriented, it is distributed to visualize the structure and the sizes of the files in your hard-drive, specifically. In accordance, users do not have to discover that the tool is good at doing this; instead, when they have this specific problem at hand, they are guided to SequoiaView, a more natural process.
	
	In designing systems, an alternative course of action would be to initially present the user with a set of patterns to choose from. After that, she/he would be offered a set of visualization techniques that better suit the pattern they are seeking for. Users may know what to look for by means of previous knowledge of the data domain, by means of known problems to be solved, by means of suspicious clues perceived along the data usage, and also by means of previous visual exploration of the data.
	
	
	\subsection{Cognitive Decision Support}
	\label{subsec:decisionsupport}
	
	\noindent{After producing patterns, vision is no longer an active agent, neither pre-attentively nor attentively. Now, the analysis follows the widely accepted pattern-then-cognition process \citep{Margolis90} \citep{MacEachren04} to achieve decision support. Indeed, according to the analytical reasoning of \cite{Thomas2005}, Information Visualization cannot ultimately provide decision support, which can only be achieved after interpreting the patterns in light of the data domain -- rightmost side of Figure \ref{fig:VisualExpressionProcess}. That is, even though users can come up with patterns without considering the underlying data, these patterns are not of great use if the domain is not deeply understood.}
	
	Insufficient domain knowledge leads to unsatisfied expectations in relation to InfoVis, creating situations in which a user is presented to a supposedly insightful visualization but, then, everything one hears is ``so what''? A disappointment that happens due to the enthusiasm according to which one can solve a wide range of problems just by looking at the data. However, visualization tools can do little if the analyst is not well-prepared to assess what the data ultimately describes and potentially carries within.

\subsubsection*{Discussion}	
\begin{figure*}[htb]
		\centering
		\includegraphics[width=\textwidth]{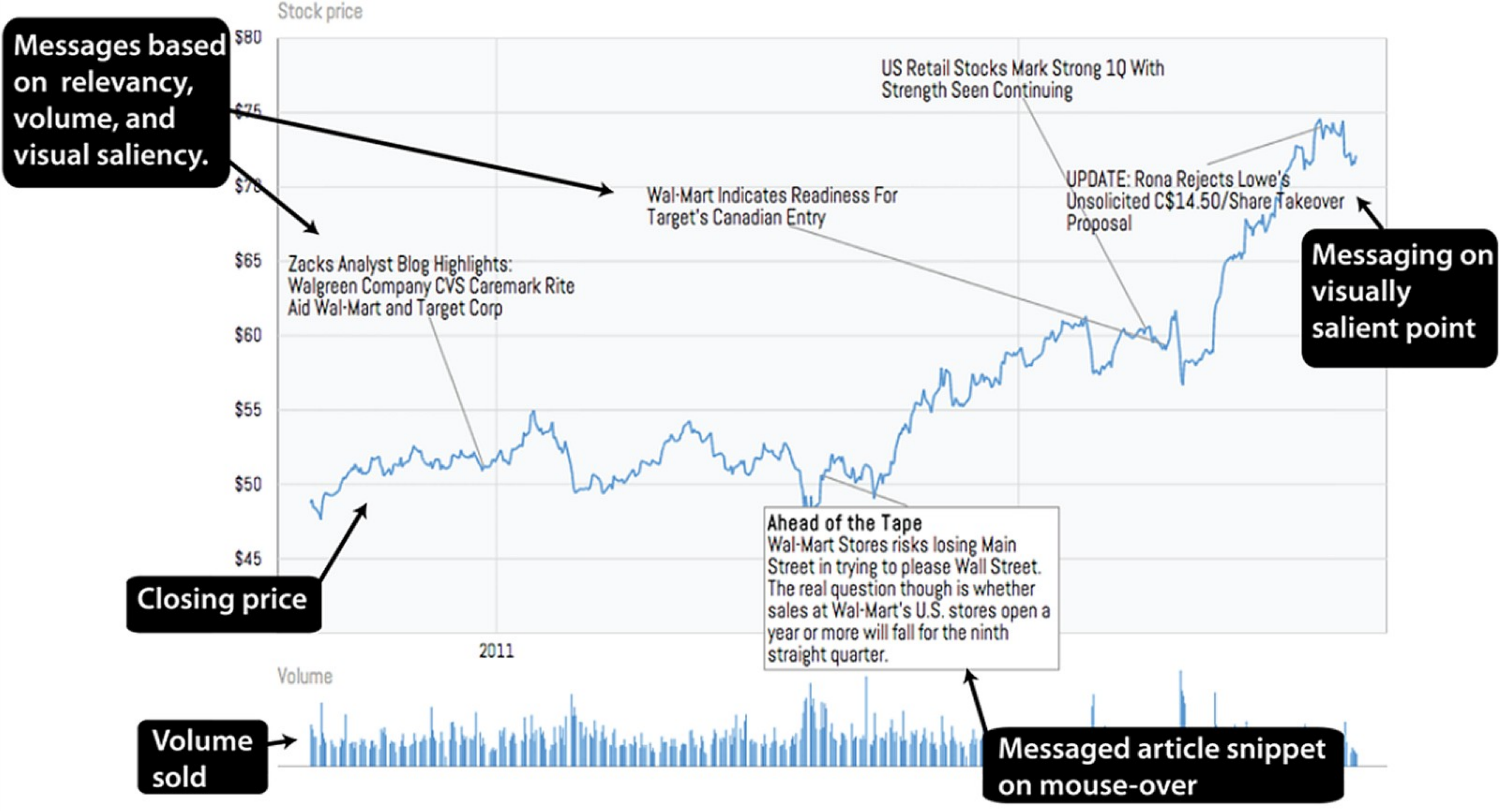}
		\caption{Example of an annotated visualization produced with Contextifier -- in the figure, labels provide domain knowledge. Reproduced from the work of \cite{Hullman2011BIV}.}
		\label{fig:hullman}
	\end{figure*}

	To prevent the aforementioned problems, InfoVis systems should define systematic means to aid the user in recording and accessing the domain knowledge related to the problems at hand. An interesting approach to overcome the gap of domain knowledge is to use annotations (or \textit{analytic provenance} \citep{Xu2015}), either automatic or manual. As pointed out by \cite{Hullman2011BIV}, annotations help to direct the user's attention and foreground particular insights, supporting the most efficient inferences. The work of \cite{Hullman2013} exemplifies this issue. Their work focuses on stock-price time series, which are hard to understand if one is not aware of the facts that influenced the behavior of the market. Their system solves this problem by identifying news that happened contemporaneously to outstanding patterns found in the plots, presenting them on demand -- illustrated in Figure \ref{fig:hullman}. Similar approaches \citep{2319219,16081474} have been proposed for genomic data, an extreme case in which domain knowledge is necessary, otherwise no reading of the data (either visual or textual) will make sense.
	
	An alternative is to use technique {\it storytelling}, as surveyed by \cite{Segel2010}. \cite{Plaisant200553} defends that advanced interfaces need to address the long-term process of analysis that may require annotation, history keeping, collaboration with peers, and the dissemination of results and procedures used. Storytelling not only attacks the problem of lacking domain knowledge, it also provides knowledge about new findings in the form of further ``story  chapters'' interactively created. The visual analysis, potentially, becomes an incremental set of insights from multiple experts in the form of bookmarks, keyword tags, text comments, and audio annotations.
	
	By bringing existing (factual) or produced (annotated) domain knowledge to the stage, abstract patterns can support reasoning by supporting conditional inferences \citep{johnson2002conditionals}, possibly influenced by statistical regularities \citep{patterson2012training}. The result is the provision of inferences for decision making, notably, predictions, alternatives, and evaluations as depicted in Figure \ref{fig:VisualExpressionProcess}.
	
	\subsection{Remarks}
	\label{subsec:discussion}
	
	In Sections \ref{subsec:pre_attentive_stimuli} through \ref{subsec:decisionsupport}, we have drawn conclusions by surveying accepted concepts of vision and cognition. Notwithstanding, we state these conclusions as conjectures with theoretical evidence and with examples. This is because the validation of these hypotheses would encompass vast experimentation; producing material enough to spam a few papers or a book, to be conservative. Therefore, we leave our discussions both as contributions -- to guide new systematizations; and, as future work -- to drive further refinements and discoveries.
	
	Our conclusions also point to a challenging systematization. Rendering all the recommended aspects, together with multiple techniques and data domains, might involve a development effort similar to that of huge software pieces, as office suites for instance. This is academically non-attractive, and economically risky. Possibly, the solution is to set a well-defined development framework for collaborative work, with ample acceptation, and standard interfacing. In the realm of machine learning, software Weka \citep{Hall2009} has achieved great success in a similar endeavor. However, the graphical nature of InfoVis, and its data preprocessing techniques, imposes big challenges.

	\section{Conclusions}
	\label{sec:conclusions}
	
	\noindent{We reviewed concepts on vision, cognition, and Information Visualization by introducing the organizational theorization named {\it Visual Expression Process}, which proposes a course of action to explain visual data analysis. Over an extensive literature survey, the theorization provides comprehension and science for data graphical presentation. It proposes a new perspective to discuss and characterize techniques. From this perspective, for instance, it would be possible to ``dissect'' a given technique in terms of the channels, the analytical perceptions and the patterns that it supports, potentially revealing strengths and weaknesses. Our contributions are as follows:}
	
	\begin{itemize}
		\item{Theoretical compendium: we plotted the Visual Expression Process to interrelate vision, cognition, and Information Visualization;}
		\item{Discussions: we provided an extensive survey from different fields of science to serve as basis for relevant debate;}
		\item{Reflections: we revisited design practices supported by examples and study cases.}	
	\end{itemize}

	Overall, we have put together key concepts to introduce an insightful consideration of visualizations. The reductionist perspective of our organization translates non-familiar concepts of vision and cognition into their corresponding effects with respect to design. In the form of a vocabulary taxonomically organized, we proposed a simplified comprehension of the factors that define techniques and systems. With our contributions, we expect to foster a more comprehensive, accessible, and applied science of visualization.
	
	
	\bibliographystyle{agsm}

\end{document}